Title: **Mixed genetic background better recapitulates developmental and psychiatric phenotypes and heterogeneity than inbred C57BL/6J mice**

Ana Dudas[1], Ana Novak[2], Caroline Gora[1], Emmanuel Pecnard[1], Nicolas Azzopardi[1], Séverine Morisset-Lopez[2], Lucie Pellissier[1*]

[1]INRAE, CNRS, Université de Tours, PRC, 37380, Nouzilly, France

[2]Center for Molecular Biophysics-CNRS UPR 4301, Rue Charles Sadron, Orléans, France

[*]Corresponding author: Lucie P. Pellissier, PhD, Team biology of GPCR Signaling systems (BIOS), INRAE, CNRS, Université de Tours, PRC, 37380, Nouzilly, France. Phone: +33 4 47 42 79 62. Email: lucie.pellissier@inrae.fr

Word count: **2392 words**

**ABSTRACT**

Preclinical models of neurodevelopmental and psychiatric conditions often rely on inbred mouse strains like C57BL/6J (B6), which exhibit limited genetic and behavioral variability. This limitation hampers the modeling of phenotypic heterogeneity, characteristic of these conditions. Recent efforts have explored the use of multiple genetically diverse hybrid strains to address this. In this study, we examined whether one single mixed genetic mouse background, C57BL/6J;129S2/SvPas (B6;129), could simultaneously recapitulate behavioral variability and display improved phenotypes relevant to neurodevelopmental and psychiatric conditions. Compared to the inbred B6 strain, mixed B6;129 mice displayed enhanced sociability and self-grooming, two key discriminating parameters between the two mouse backgrounds and core features of such conditions, alongside a broader spectrum of individual behavioral variability. Although overall behavioral variability was comparable across backgrounds, B6;129 mice were less susceptible to the effects of chronic social isolation than their B6 counterparts.

Together, these findings support that the B6;129 mixed background offers a more representative model of individual variability and behavioral traits associated with neurodevelopmental and psychiatric conditions, thereby enhancing its translational relevance in preclinical studies.

**INTRODUCTION**

Neurodevelopmental and psychiatric conditions, such as schizophrenia or autism spectrum disorder (ASD), are defined by the Diagnostic and Statistical Manual of Mental Disorders[1], based on the behavioral criteria, with deficits in social interaction being a common feature. These conditions are widely heterogeneous across individuals, driven by complex interactions between genetic predispositions and environmental factors, which together shape neural circuitry and behavioral outcomes[2–4].

Over the past two decades, mice have served as valuable models for studying those conditions and testing new treatments, as they can exhibit both impaired social behaviors and co-occurring or negative features[5]. Inbred strains, created by mating brother-sister pairs over at least 20 generations to achieve ~99% homozygosity, are widely used in neurological research to examine genetic and environmental mechanisms and the effect of potential treatments. Among these, the C57BL/6 (B6) is the most commonly used strain due to its reduced genetic variability and well-characterized genome[6]. B6 mice display average sociability relative to other inbred strains[7,8]. Several inbred strains have even been identified as mouse models of ASD. Notably, BTBR T$^+$ tf/J mice exhibit profound social and communication impairments, as well as increased repetitive and stereotyped behaviors compared to B6 mice[8–12]. In contrast, FVB/NJ mice show high sociability, often serving as a control to low-sociability strains in experimental assays[7,8,12]. The 129Sv (129) strains, commonly used for generating transgenic models, display variable sociability levels, mirroring the diversity observed in human conditions[8,13].

While inbred mice are valuable for studying the effects of single-gene deletions or environmental factors independently, findings derived from these homogeneous backgrounds may not generalize to genetically heterogeneous conditions[14–16]. To address this limitation, genetically diverse mouse populations have been developed to model complex gene–environment interactions affecting behavioral traits, neurodevelopmental and psychiatric conditions. Outbred strains (e.g., Swiss, CD1) are maintained by random mating in large colonies to preserve genetic diversity. Additionally, inbred mouse lines such as the BXD or Collaborative Cross (CC), derived from controlled crossing of inbred strains and/or wild-derived mice, offer genetic diversity with fully sequenced genomes[17–20]. Notably, wild-derived mice often exhibit behavioral traits distinct from lab inbred mice; for instance, female wild-derived mice maintain a social hierarchy lost in laboratory strains[21]. A recent study revealed that crossing B6 heterozygous females for the high-confidence ASD risk gene ($Cdh8^{+/-}$) with 27 reference CC males, used to model human genetic heterogeneity, strongly influences behavioral phenotypes[22].

F1 hybrids, the direct offspring of two different inbred strains strictly heterozygous at all loci, generally exhibit enhanced physiological robustness compared to inbred lines[23]. However, behavioral studies in hybrid genetically homogeneous mice have focused on memory and anxious-like behaviors[6,23]. In contrast, mixed background mice, referring to F2 or later-generation crosses of two inbred strains without intentional genetical selection carry a unique, heterogeneous combination of parental alleles. These mixed background mice may better model human genetic diversity, particularly in neurodevelopmental and psychiatric conditions that are influenced by multiple genetic and environmental risk factors[24,25]. To our knowledge, the impact of using mixed background mice versus standard inbred strains for assessing psychiatric-related conditions has not been investigated. In this study, we compared behavioral phenotypes relevant to these conditions between mixed C57BL/6J;129S2/SvPas (B6;129) background and B6 inbred mice. We evaluated behavioral variability and explored how an environmental model manifests across these genetic backgrounds.

## RESULTS

**Mixed background mice exhibit enhanced sociability**

While the B6 is the most commonly used strain in research due to its low variability[6], we investigated whether introducing genetic diversity by mixing B6 with the 129 inbred strain, known for comparable sociability but greater behavioral variability[8,13], would alter sociability. Therefore, to assess social interaction and its potential variability, we compared inbred B6 mice with mixed background B6;129 mice using standardized social behavior tests, including interactions with either a familiar (cage mate) or an unfamiliar conspecific.

Strikingly, although both backgrounds were sociable, B6;129 mice showed increased social interaction, particularly with unfamiliar mice, compared to B6 mice in the sociability and social novelty phases of the three-chambered test, as well as in the reciprocal interaction assay (**Figure 1A-E**, **Table S1**). Notably, the higher inter-individual variability observed in B6;129 mice was specific to unfamiliar interactions, as no differences were found in familiar interactions. Moreover, in B6;129 mice, interactions with unfamiliar conspecifics were positively correlated across both tests, suggesting stable social behavior traits across tasks and days (**Figure 1F**). No significant sex differences were detected across backgrounds (**Table S1**).

To exclude the possibility that differences in social interaction stemmed from altered social memory or olfactory function, we assessed the performance of both backgrounds using neutral and social odors, as well as subsequent exposure to unfamiliar and familiar mice. B6 and B6;129 mice performed comparably, with the exception again of reduced interaction with unfamiliar animals in trial 4 (**Figure 1G-H**), indicating that both strains exhibit similar social recognition and olfactory processing.

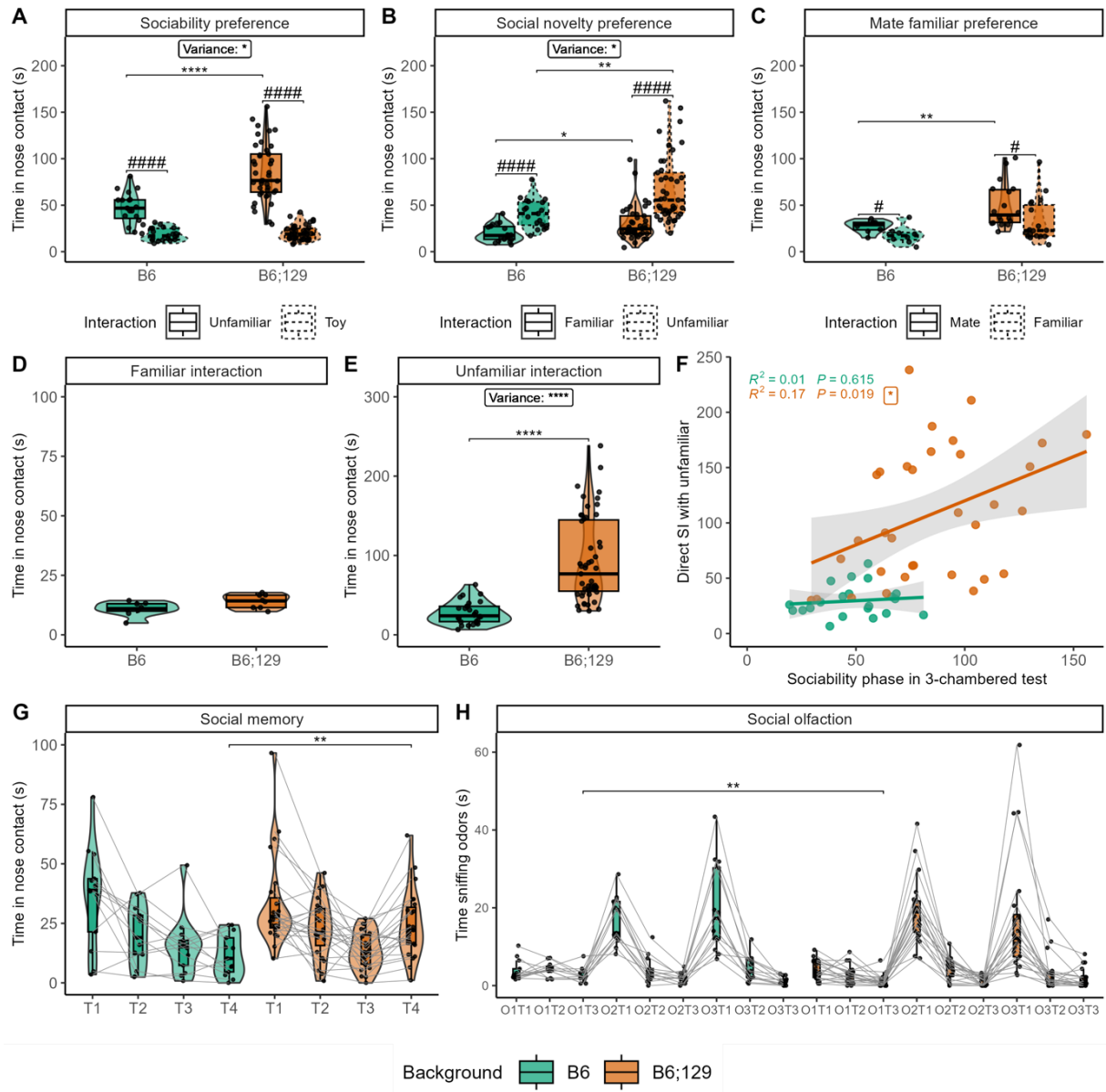

*Figure 1. Mixed B6;129 mice exhibit enhanced sociability and variability with unfamiliar mice.* In the 10 min three phases of the three-chambered test, B6;129 mice (orange) showed enhanced sociability (unfamiliar WT mouse 1 vs object; A), social novelty preference (familiar WT mouse 1 vs novel unfamiliar WT mouse 2; B) and familiar cage mate (familiar cage mates vs familiar WT mouse 2; C) preference compared to B6 mice (green). In the 10 min reciprocal social interaction, B6;129 mice showed enhanced social interaction with a background-, sex- and age-matched unfamiliar mouse (D) but not familiar cage mates (E) compared to B6 mice. Nose contact with an unfamiliar mouse in the reciprocal social interaction was positively correlated with those with an unfamiliar mouse in the sociability phase of the three-chambered test (F). In the 5 min social memory test (G), both B6;129 and B6 mice exhibited similar social recognition and habituation to the WT interactors over the four trials. In the odor habituation-dishabituation assay (H), B6;129 and B6 mice displayed similar neutral (O1, blossom flower) and social odor (O2 sex-matched or O3 opposite sex WT urines) recognition. Data are presented as individual data, mean ± sd (statistics, n and sex ratio in Table S1). Groups were compared by Kruskal-Wallis tests followed by Dunn post hoc tests, with stars indicating background effect and hash chamber effect (P = adjusted p-value), while inter-individual variability was compared by Levene's variance test. * or # $p < 0.05$; ** or ## $p < 0.01$; *** or ### $p < 0.001$; **** or #### $p < 0.0001$.

To further explore social interaction in a more complex and dynamic setting, we used the Live Mouse Tracker (LMT) system[26] to expose B6 and B6;129 mice to either familiar cage mates or a mixed group of four unfamiliar B6 and B6;129 mice over consecutive days (**Figure 2**, **Table S2**). B6;129 mice displayed increased time spent following both familiar and unfamiliar mice, compared to B6 mice (**Figure 2A, C**). This increased behavior was not accompanied by elevated oral–anogenital sniffing, a behavior often linked to aversive or aggressive social interactions, particularly in males[27,28], suggesting that the enhanced following behavior in B6;129 mice reflect increased social motivation and interest rather than aggression. Sex-specific differences emerged more clearly in the B6;129 background. Female B6;129 mice exhibited increased huddling and social approach behaviors toward familiar animals. For nose-to-nose contacts, sex effect was observed in both backgrounds but was more pronounced in B6;129 mice (**Figure 2B**, **Table S2**), indicating that the mixed background enhanced the detection of sex-dependent behavioral differences.

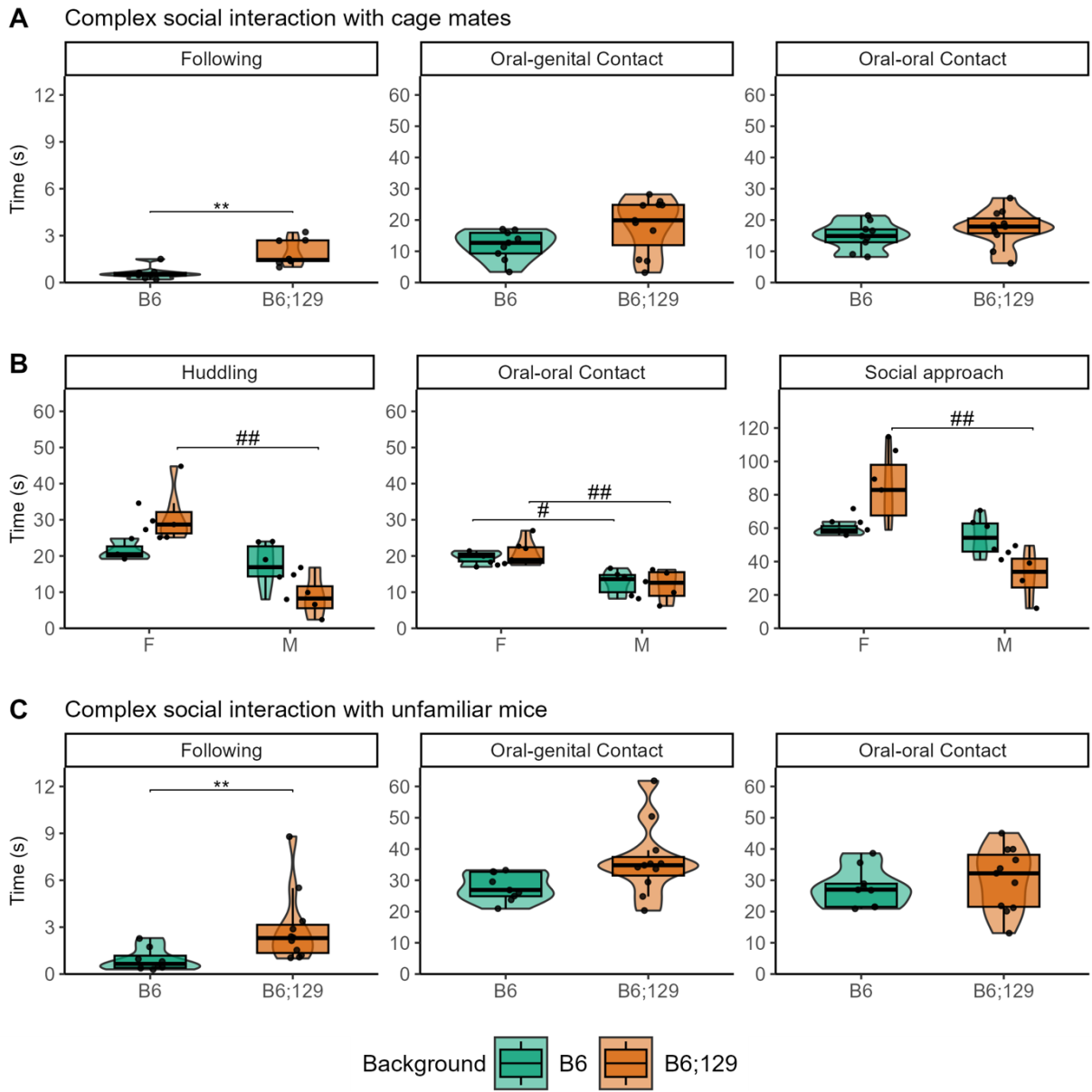

*Figure 2. Mixed B6;129 mice show improved social motivation and sex differences in the Live Mouse Tracker.* In the 10 min complex social interaction in the Live Mouse Tracker, B6 mice (green) and B6;129 mice (orange) displayed increased time following familiar cage mates (A-B) and in the subsequent trial with a mix of 2 unfamiliar B6 mice and 2 unfamiliar B6;129 mice while no differences in nose contacts were observed. Sex differences were identified when interacting with familiar cage mates (B), with females showing increased time huddling and in social approach compared to males in the B6;129 background, as well as nose-to-nose contacts in both backgrounds. Data are presented as individual data, mean ± sd (statistics, n and sex ratio in Table S2). Groups were compared by Kruskal-Wallis tests followed by Dunn post hoc tests, with stars indicating background effect and hash sex effect (P = adjusted p-value), while inter-individual variability was compared by Levene's variance test. * or # $p < 0.05$; ** or ## $p < 0.01$; *** or ### $p < 0.001$.

In conclusion, mixed B6;129 background mice display enhanced social interaction toward unfamiliar conspecifics, accompanied by increased variability and sensitivity to sex differences.

**Mixed background mice display increased self-grooming and reduced anxious-like behaviors**

We next assessed whether genetic background influenced stereotyped, compulsive, or restrained behaviors, as well as anxious-like behavior, cognitive flexibility, and locomotion (**Figures 3-4**, **Table S1**).

B6;129 mice exhibited increased time in self-grooming, but no differences in the number of head shakes (repetitive and stereotyped behaviors, respectively), while both variances were significantly higher compared to B6 mice (**Figure 3A**). In contrast, B6 mice displayed increased number of digging episodes (but not the time), indicative of natural foraging or heightened anxious-like behaviors[29]. However, no differences were observed in marble burying (obsessive-compulsive–like behavior) and Y-maze spontaneous alternation (restrained behavior and cognitive flexibility) between backgrounds (**Figure 3B-C**), suggesting that the increased digging in B6 mice likely reflects elevated anxious-like behavior rather than compulsivity.

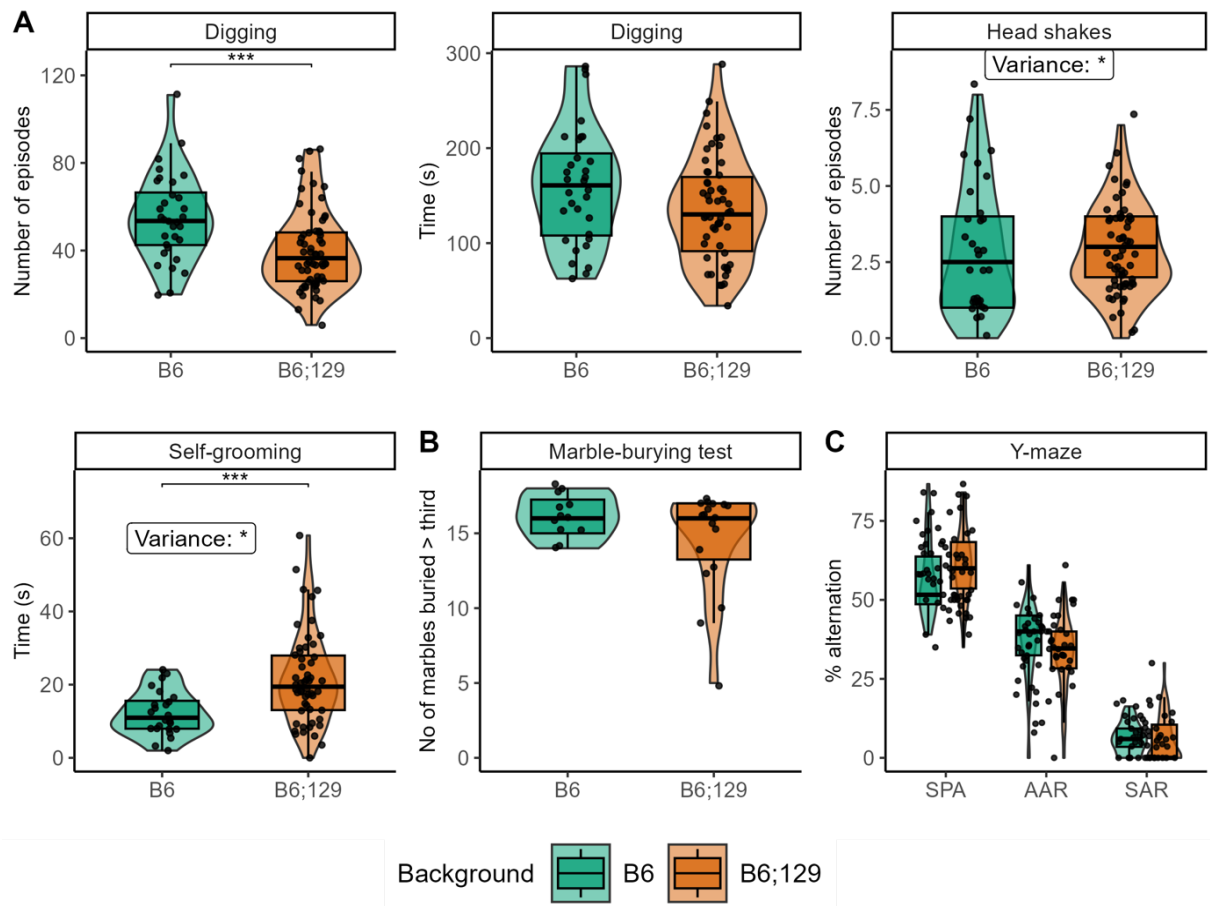

*Figure 3. Mixed B6;129 background mice display increased self-grooming.* In the 10 min motor stereotypies (A), B6 mice (green) displayed enhanced number of digging episodes, and no difference in the time spent digging or head shake episodes, while B6;129 mice (orange) showed increased time in self-grooming. No difference in the marble burying test (B) or in the Y maze alternation pattern (C) was observed between B6 mice and B6;129 mice. Data are presented as individual data, mean ± sd (statistics, n and sex ratio in Table S1). Groups were compared by Kruskal-Wallis tests followed by Dunn post hoc tests, with stars indicating background effect (P = adjusted p-value), while inter-individual variability was compared by Levene's variance test. * p < 0.05; ** p < 0.01; *** p < 0.001.

Regarding other behaviors, no significant differences were found between backgrounds in spatial memory, novel object recognition, or locomotor activity (**Figure 4A-E**, **Table S1**). However, anxious-like behavior was significantly higher in B6 mice compared to B6;129 in the novelty-suppressed feeding test, whereas no differences were observed in the elevated plus maze or in the time spent in the center of the open field (**Figure 4D-F**), supporting the anxiety-related behaviors detected via digging behavior. No significant sex differences were detected across these tests (**Table S1**).

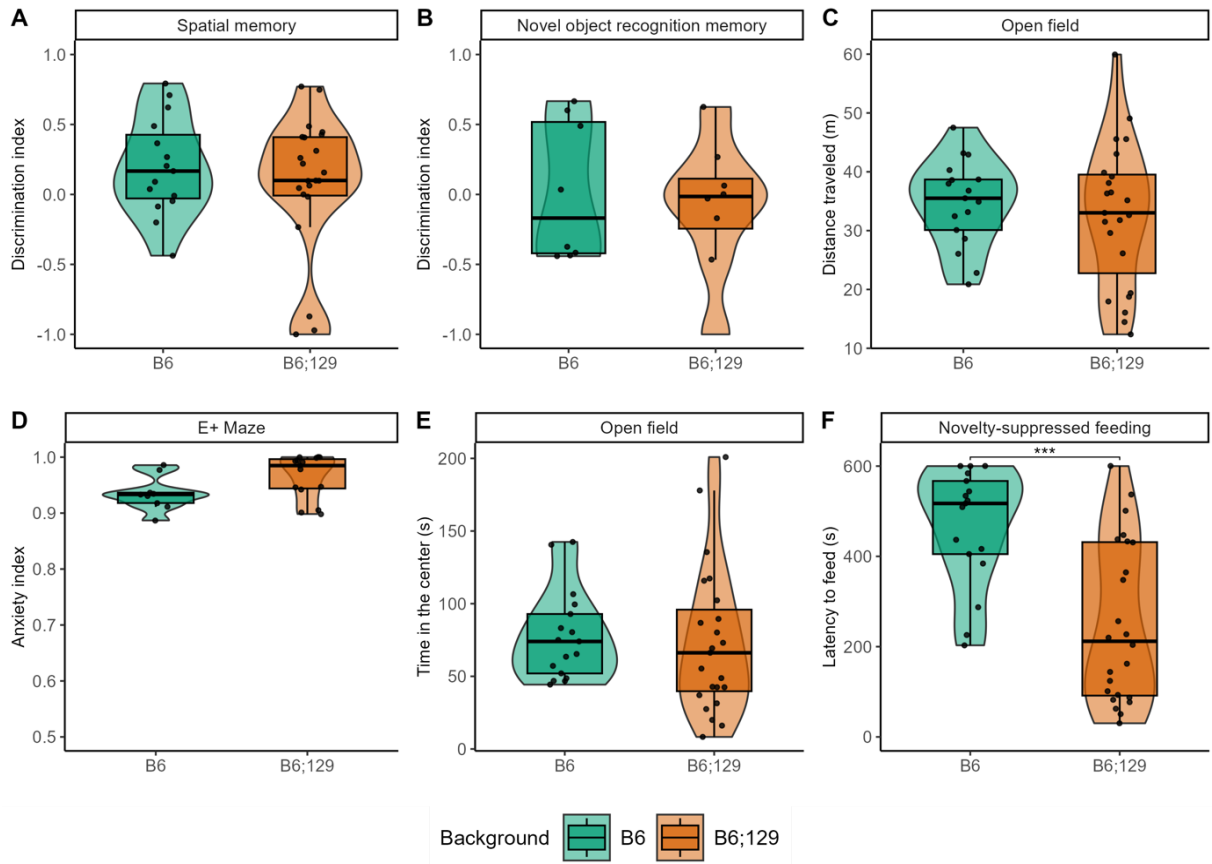

*Figure 4. Inbred B6 mice display increased anxious-like behaviors. In the spatial object location (A) and novel object (B) recognition memory tests, B6 mice (green) and B6;129 mice (orange) performed similarly, as well travelled the same distance in the open field (C) for 10 min. No difference in anxious-like behavior was observed in the elevated plus maze (D) or time in the center in the open field arena (E) for both backgrounds, while B6 mice showed increased latency to feed in the novelty-suppressed feeding tests (F), compared to B6;129 mice. Data are presented as individual data, mean ± sd (statistics, n and sex ratio in Table S1). Groups were compared by Kruskal-Wallis tests followed by Dunn post hoc tests, with stars indicating background effect (P = adjusted p-value), while inter-individual variability was compared by Levene's variance test. \* p < 0.05; \*\* p < 0.01; \*\*\* p < 0.001.*

In conclusion, B6;129 mice displayed increased self-grooming and reduced anxious-like behaviors, compared to B6 mice. Unlike social interaction parameters, behavioral variability in the mixed background was less pronounced for these behaviors.

**Sociability and motor stereotypies are key discriminating parameters between the two backgrounds**

To identify the key parameters that robustly discriminated between the two backgrounds, we conducted a principal component analysis (PCA) using all behavioral parameters (**Figure 5**, **Tables S1-2**).

Only PC1 (Dim1) significantly differed between B6 and B6;129 mice and was driven primarily by enhanced sociability and social novelty in the three-chambered test and self-grooming behavior in B6;129 mice, as well as increased digging behavior in B6 mice (**Figure 5A**). Although less pronounced than in classical tests, PCA analysis applied to the LMT dataset revealed that the top five discriminating factors were social parameters (**Figure 5B**). PC2 significantly distinguished the backgrounds, with B6 mice characterized by increased isolation, and B6;129 mice by higher contacts.

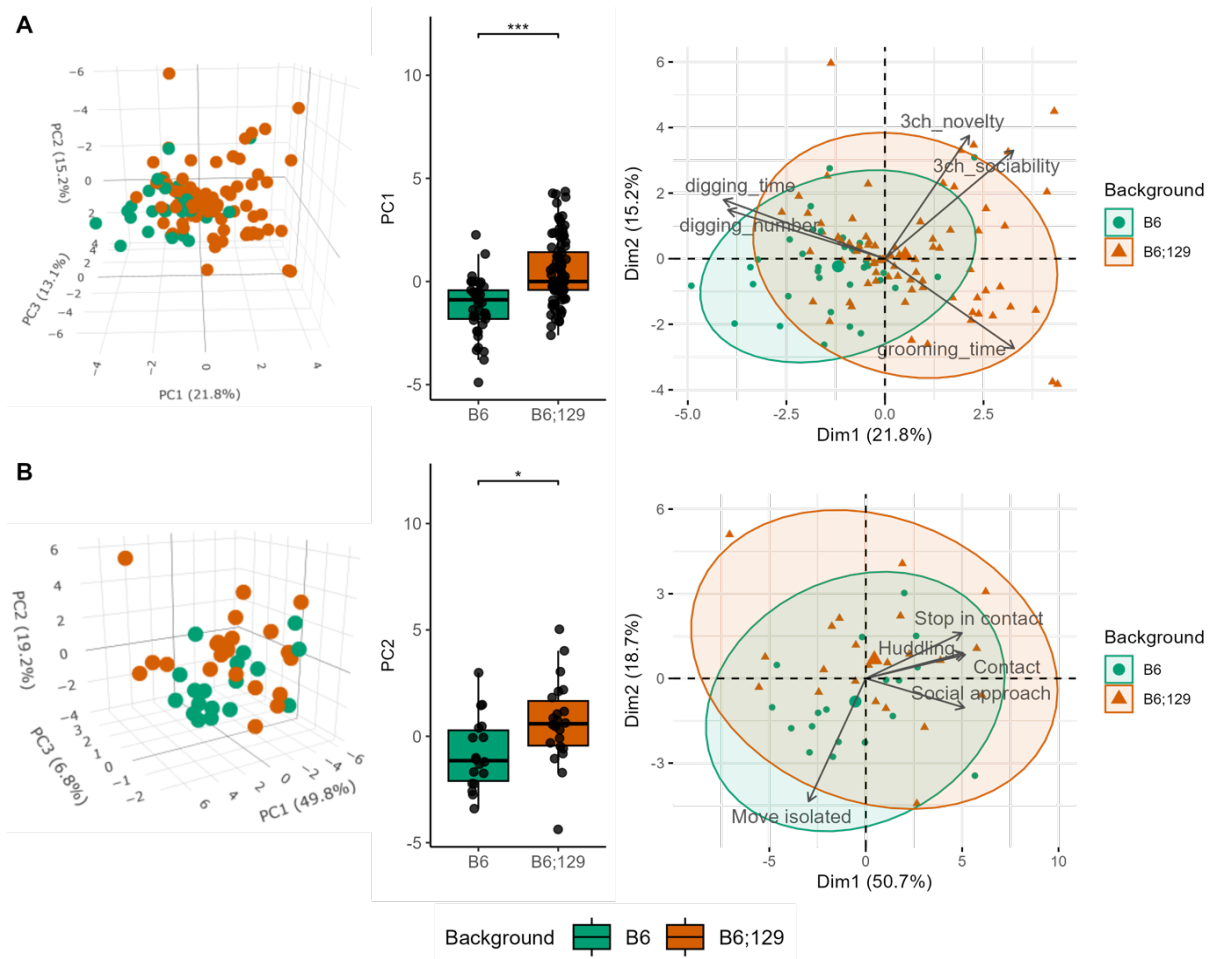

*Figure 5. Sociability and motor stereotypies discriminate between backgrounds. Based on the behavioral phenotypes of B6 mice (green) and B6;129 mice (orange) in the reciprocal social interaction, three-chambered, motor stereotypies, Y-maze, open field, and novelty-suppressed feeding tests (A), principal component analysis revealed higher variability in B6;129 background in the first principal component (left), confirmed by a significant increase of the scores for the first principal component in B6;129 compared to B6 background (center). The most contributing behavioral parameters to this variability in B6;129 background were sociability and stereotyped behavior parameters (right). Based on the behavioral phenotypes of B6 (green) and B6;129 mice (orange) in the Live Mouse Tracker interacting with cage mates or unfamiliar age- and sex-matched B6 and B6;129 mice (B), the principal component analysis revealed increased variability of the B6;129 mice in the second principal component (left), evidenced by an increase in the scores for the second component (center). The most discriminating parameters contributing to this variability in the B6;129 background were social interaction parameters (right). Data are presented as individual data, mean ± sd (Tables S1-2). Groups were compared by the one-way ANOVA followed by Tukey's posthoc tests, with stars indicating the background effect. * p < 0.05; ** p < 0.01; *** p < 0.001.*

In conclusion, these analyses demonstrate that sociability and motor stereotypies, both core ASD-like traits, are the key discriminating behavioral parameters between the inbred and mixed genetic backgrounds.

**Mixed background is less sensitive to chronic social isolation, a mouse model of neurodevelopmental and psychiatric conditions**

To assess whether the mixed B6;129 background better models behavioral heterogeneity in a mouse model of ASD and schizophrenia[30–33], or whether findings are specific to the inbred B6 background, we compared the effects of four weeks post-weaning chronic social isolation in mice (isolated) in both genetic backgrounds (**Figure 6**, **Table S3**).

B6;129 wild-type mice raised in groups (WT) and isolated mice showed enhanced social interaction with an unfamiliar conspecific across multiple social assays (**Figure 6A-B**). These groups also displayed greater inter-individual variability compared to B6 WT mice (**Table S3**), confirming our previous baseline observations between backgrounds. Chronic social isolation increased social interaction with an unfamiliar, isolated conspecific in the reciprocal interaction test, an effect that was significant only in the B6 background, but did not change social motivation in the three-chamber test when interacting with an unfamiliar WT mouse (**Figure 6A-B**). Interestingly, although sociability was more variable among isolated B6;129 mice compared to their respective WT, isolated B6;129 mice showed a stronger preference for the mouse over the object than isolated B6 mice. This suggests that social interaction in the B6;129 background is less sensitive to the effects of chronic social isolation.

WT B6;129 mice also displayed increased self-grooming compared to WT B6 mice (**Figure 6C**), confirming our results in previous cohorts. However, this difference was not observed in isolated animals. Furthermore, both B6;129 WT and isolated mice spent less time digging and displayed fewer head shakes compared to B6 mice. Although chronic isolation was not expected to affect head shakes, a proxy of stereotyped behavior[31,32], it amplified the behavioral differences between the two genetic backgrounds.

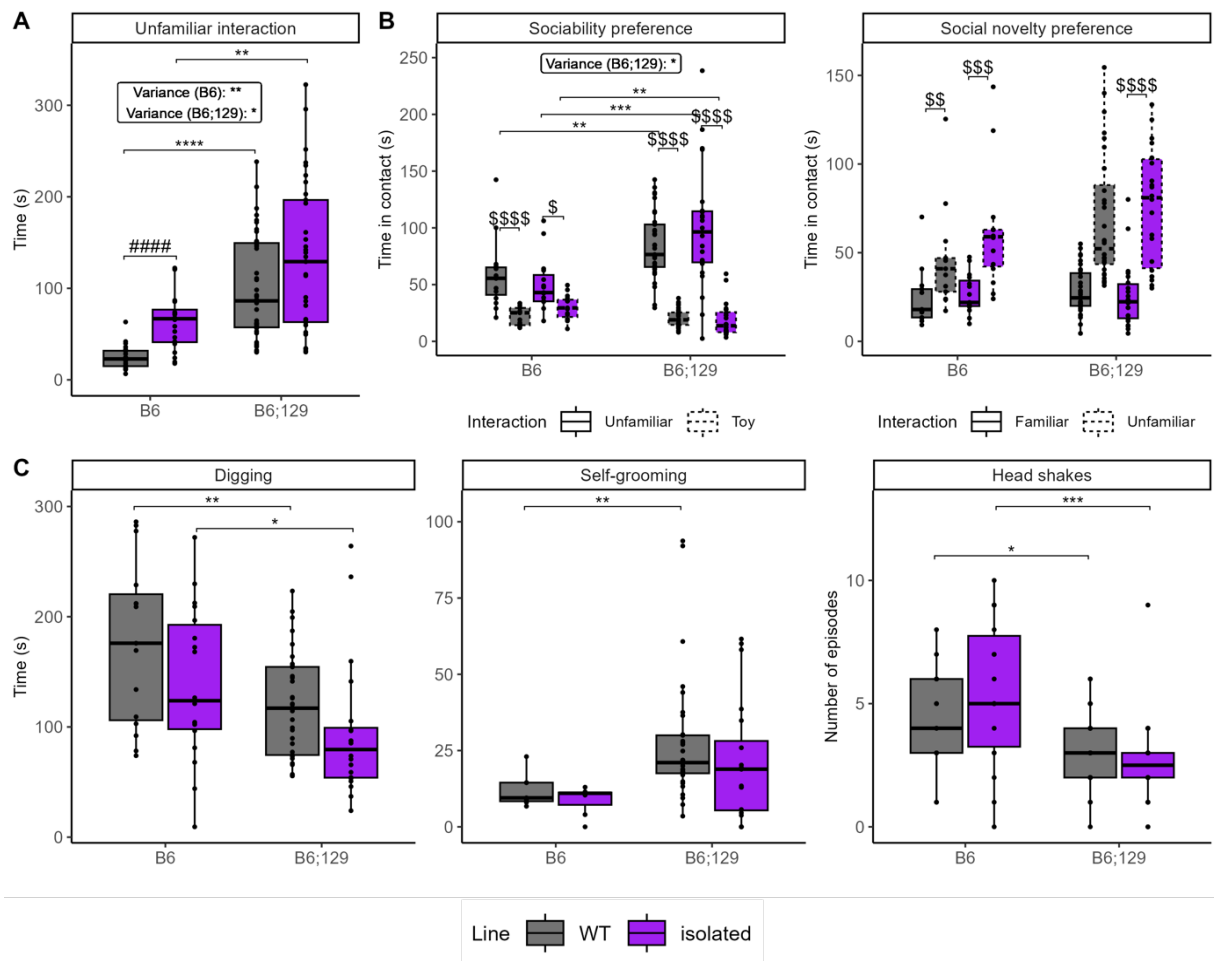

*Figure 6. Mixed B6;129 background enhances sociability and self-grooming in WT, but not in chronically isolated mice.* In the reciprocal social interaction test, chronically isolated mice (purple) revealed significant differences compared to WT controls (gray) only in the B6 background, while chronic isolation increased variance in both backgrounds. Social interaction was significantly increased in both WT control and isolated mice from the B6;129 compared to B6 background (A). In the sociability phase of the three-chambered test, sociability preference was observed for both chronically isolated mice and WT control, while this preference was significantly higher in B6;129 background with a higher variance (B). In the motor stereotypy test, there was no difference in the time spent digging or self-grooming between WT and chronically isolated mice, while both WT and isolated mice in B6 background showed increased time spent digging and number of head shakes, while increase in time spent self-grooming was observed in B6;129 WT mice compared to B6 conspecifics (C). Data are presented as individual data, mean ± sd (statistics, n and sex ratio in Table S3). Groups were compared by Kruskal-Wallis tests followed by Dunn post hoc tests, with stars indicating background effect, hash housing effect, and dollar chamber effect (P = adjusted p-value), while the inter-individual variability was compared by Levene's variance test. *, # or $ p < 0.05; **, ## or $$ p < 0.01; ***, ### or $$$ p < 0.001; ****, #### or $$$$ p < 0.0001.

In conclusion, while both backgrounds displayed greater behavioral heterogeneity following chronic social isolation, the mixed B6;129 genetic background appeared less sensitive to its effects.

**DISCUSSION**

Although more variable, as expected from their genetically heterogeneous background, mice raised in the mixed B6;129 background exhibited a robust and replicable increase in social interactions with unfamiliar conspecifics compared to those maintained in the inbred B6 background. Given that 129 display comparable sociability to B6 mice[6,8,13], the enhanced sociability and broader inter-individual variability observed in the B6;129 mixed line likely result from increased genetic diversity rather than from one parental strain dominating the phenotype. Although less pronounced than in the mixed background, this increase in sociability is consistent with findings from some hybrid CC mouse lines, which exhibit a modest increase in social interaction relative to their inbred B6 parental strain[22].

Crucially, the enhancement in social interaction in B6;129 mice cannot be attributed to differences in olfaction of social memory, as both backgrounds performed similarly in those tasks. This is important considering that olfactory processing plays a central role in rodent social interactions[34] and that olfactory deficits are a known driver of impaired social behavior in ASD mouse models (e.g., BTBR *vs.* B6)[11]. However, it remains possible that differences in early-life maternal care between strains contributed to adult behavioral outcomes. While some studies report no difference in maternal behavior[35], others describe shorter pup retrieval latency in 129 females[36], which may influence offspring social behavior.

At a molecular level, oxytocin receptor levels may help to explain the observed differences in social interaction between the two backgrounds. For example, B6 mice exhibit reduced oxytocin receptor density, particularly in the medial preoptic area (MPOA), compared to 129 mice[35]. Moreover, levels of oxytocin receptors in the nucleus accumbens have been implicated in regulating social bonding in monogamous prairie voles[37], and oxytocin receptor signaling in this brain region controls social reward and interaction with unfamiliar conspecifics in mice[38].

Oxytocin receptor levels may specifically underlie the background difference in unfamiliar, but not familiar, social interaction that we observed. Supporting this, we have previously demonstrated distinct oxytocin kinetics and social plasticity during these two different types of social interactions in B6;129 mice[30].

Similar to social interaction, we observed that WT mice on the mixed B6;129 background displayed increased self-grooming behavior, which was not found in inbred B6 and 129 lines[39]. This increased repetitive behavior was modest compared to *Shank3* KO mice in the same background[31]. In contrast, B6 mice showed heightened anxious-like behavior, consistent with previous reports comparing inbred B6 and 129 lines[6,12,35,39], which also reported increased locomotor activity, although not significant in our study. No major differences in learning, memory, or cognitive flexibility were observed between the two backgrounds, although subtle effects on long-term memory or age-related decline, as previously documented[6,23], cannot be excluded.

Notably, social interaction and motor stereotypies, two core behavioral features relevant to neurodevelopmental and psychiatric conditions, particularly ASD, emerged as the most discriminating behavioral parameters between the inbred B6 and mixed B6;129 genetic mouse backgrounds. Despite the expected behavioral variability associated with their genetic background, B6;129 mice consistently amplified sociability and repetitive behaviors, facilitating the potential detection of phenotypes in *Shank3* KO mice[31]. Interestingly, B6;129 mice were also less sensitive to social isolation, as previously reported[30], compared to the inbred B6 background, suggesting improved behavioral adaptability, along with greater physiological robustness[23]. These findings align with studies on CC and BXD lines, which demonstrated that genetic diversity can reveal or attenuate behavioral phenotypes of the inbred B6 parental strain[22,40].

Importantly, our findings address common concerns that mixed backgrounds might mask the effects of genetic manipulations due to increased variability. In fact, both genetic backgrounds showed enhanced behavioral variability following chronic isolation. Conversely, reliance on a single inbred strain may restrict generalizability: for example, only B6 mice displayed increased social interaction in response to isolation (with an isolated mouse), a pattern not significant in the B6;129 line, indicating strain-specific responses that may limit translational relevance when using inbred models alone[14,15].

Overall, our findings highlight the genetic and behavioral relevance of using the mixed B6;129 background for modeling neurodevelopmental and psychiatric conditions. Compared to the inbred B6 line, the B6;129 background better captures the behavioral heterogeneity characteristic of these conditions in humans. Incorporating genetic diversity into preclinical neurological research not only improves the reproducibility of findings but also enhances their translational value, addressing key limitations of conventional inbred models.

**METHODS**

*Animals*

We confirm that the experimental protocol was approved by Ethics Committee CEEA Val de Loire N°19 animal care committee's regulations and the French and European Directives (APAFIS #18035-2018121213436249). The experimental procedures were performed following the ARRIVE guidelines[41].

Following breeding of male or female 100% pure C57BL/6J (B6) with respective female or male 100% pure 129/S2SvPas (129) (Charles River, France), the hybrid F1 generation was backcrossed over five generations to generate the 50%-50% mixed C57BL/6J;129/S2SvPas (B6;129) background mice. The mice were housed in the same room of a conventional animal facility under 12-hour light/dark cycle, with food and water *ad libitum*, controlled temperature (21°C), and humidity (50%), in social groups of 2-4 sex- and genotype-matched animals per type 2 conventional cage or isolated for four weeks as previously done[31] prior to behavioral phenotyping. All mice used in the experiments were two-month-old sexually naïve males and females at the start of the behavioral experiments from 2-4 distinct cohorts.

*Behaviors*

All behavioral tests were conducted at the start of the light phase in a dedicated quiet room, with one assay per day under a dim light intensity of 15 lux (unless stated otherwise) using the standard behavioral equipment, recorded and analyzed *a posteriori*, as previously described[30,31]. The experiments were performed by a trained experimenter, mostly female, who was not blinded to the background or housing conditions. Parameters in the open field, Y-maze, elevated plus maze, and habituation in the three-chambered test were automatically analyzed using the animal tracking ANY-maze software (Stoelting, Ireland), while complex social interactions were automatically scored using the Live Mouse Tracker (LMT)[26]. Motor stereotypies, olfaction, latency to feed in the novelty-suppressed feeding, and social interaction tests were manually scored *a posteriori* by a trained experimenter blinded to the conditions using BORIS Software[42] or directly for the number of buried marbles in the marble-burying assay.

*Sociability and social memory tests*

In the reciprocal social interaction test, mice interacted with a sex-, background-, and genotype-matched cage mate or an unfamiliar mouse for 10 min in the open field arena (40 x 40 cm)[30]. In the three-chambered test, experimental animals were exposed to subsequent habituation (empty *vs* empty cylinders located in the opposite chambers), sociability (sex-matched unfamiliar WT mouse 1 (trained twice over 20 minutes in a cylinder prior to testing) *vs* Lego Duplo toy), social novelty (familiar mouse 1 vs a novel unfamiliar WT trained mouse 2 replacing the toy), and mate preference (familiar mouse 2 vs a cage mate replacing mouse 1) phases on the same day[30]. Complex social interactions in the 50 x 50 cm LMT arena[26,31] were examined across distinct and consecutive trials with up to 4 familiar cage mates or a mix of unfamiliar age- and sex-matched B6 and B6;129 mice. In the social recognition habituation-dishabituation memory test, after 5 min habituation in a conventional cage with clean bedding, experimental mice were subjected to three consecutive 5-min trials with the same unfamiliar age-, sex-matched B6;129 WT mouse, separated by one and 24-hour intervals, followed by a fourth trial with a novel unfamiliar B6;129 WT mouse after one-hour interval. "Nose contact" was defined as oral-oral and oral-genital contacts, unless specified.

*Olfactory habituation/dishabituation test*

After 5 min habituation in a conventional cage with clean bedding and a clean 2 cm x 2 cm Whatman paper tapped to the cage wall at the height of 4 cm (later soaked with 20 µL of each odor and replaced by a clean paper between each trial), sniffing (<1cm) of the soaked paper was measured by exposing individual mouse to three 2-min trials of neutral blossom flower odor (diluted 1:2 in saline), followed by three trials of same-sex urine and three trials of opposite-sex urine, with a one-minute inter-trial interval. Urine was collected from ten B6;129 WT males or females, not included in the study.

*Compulsive, repetitive, and stereotyped behaviors*

Repetitive behavior (self-grooming and digging) and motor stereotypies (vertical jump, circling, head shakes, and scratching), as well as immobility and rearing (spatial exploration), were measured for each mouse placed in a clear and clean conventional cage filled with a 3-4 cm thick layer of fresh litter for 10 min at 400 lux. Compulsive/anxious-like behavior was assessed in the marble-burying test by measuring the number of marbles buried more than two-thirds by each mouse exposed to 20 glass marbles (1.5 cm diameter) equally spaced on 4 cm thick fresh litter in a clean conventional cage closed with a transparent plexiglass for 15 min. To assess perseverative behavior, the percentage of spontaneous alternation (SPA), alternative arm returns (AAR), and same arm returns (SAR; perseverative behavior) was determined from a triplet of arm entries for each mouse placed in a (35 x 5 x 10 cm) Y-maze test for 5 min.

*Locomotion and anxious-like behavior*

Locomotion (total distance traveled and mean speed) and time in the center were evaluated for each mouse in the open field arena for 10 min. In the novelty-suppressed feeding test, the latency to reach three pellets of ordinary lab chow placed on a white tissue in the center of the open field arena filled with a thin layer of fresh litter was measured for each 16h-food-deprived mouse exposed to 60 lux. In the elevated plus maze test (4 arms, 5 cm wide and located 80 cm above the ground), the number and time spent in the opened or closed arms, the anxiety index measured as the time spent in the closed arm / total time in both arms, as well as distance traveled and immobility for each mouse, were evaluated for 5 min.

*Spatial location and novel object recognition tests*

In the spatial object location recognition test, after two habituations for 10 min with two Lego Duplo toys located on the same side of the center border in the open field area with 1.5h intervals, spatial memory of the mice was tested for 10 min by placing one Duplo ("spatial object") to the diagonal side of the arena while the other one remained at the same position ("old object"). In the novel object recognition test, after the same two habituations, mice were tested for 10 min 24 hours later for mid-term object recognition memory by replacing one toy by a bigger Lego Duplo toy with a different color ("novel" object) and, after 1.5h interval for short-term recognition memory by replacing the small habituation Duplo by a Falcon tube (second "novel" object). The discrimination index was calculated as the time spent with the spatial or novel object – old object / total time with the objects.

***Experimental design and statistical analysis***

Power analysis of animal sample size calculated a minimum of 8 mice per group (power of 80%, alpha level of 5% ($p = 0.05$), differences in means of 41%). The sample size was greater than 8 animals per group, as all animals within a cage were tested to avoid any bias due to dominance/subordinate effect and potential death of animals. The exact sample sizes and sex ratio are provided in **Tables S1-3**.

Criteria to exclude animals from behavioral tests were over 30% of the time spent immobile and/or a minimum of two explorations of all arms or compartments. The animal outliers (±3 standard deviations) detected in each behavioral assay were removed. Therefore, a total of 14 animals (11 B6;129 and 3 B6 mice) were excluded from the analysis. All data and statistical analyses were performed using R software (version 4.3.1). For comparisons between groups, the non-parametric Kruskal-Wallis tests with Dunn's posthoc tests were performed using the rstatix package[43]. P-values were adjusted with Benjamini-Hochberg correction[44]. The correlation of social interaction parameters from the reciprocal social interaction test and the sociability phase of the three-chambered test was assessed using the ggpmisc package[45]. Principal component analysis (PCA) was performed using the FactoMineR package[46] to visualize the inter-individual variability in the reciprocal social interaction, three-chambered, motor stereotypies, Y-maze, open field, and novelty-suppressed feeding tests in each background. Missing values (NAs) in the behavioral dataset used in PCA were imputed by replacing each missing entry with the mean value of the corresponding variable across all samples. This was done using a column-wise operation where, for each variable, all missing values were substituted by the variable's calculated mean, excluding NAs. To assess whether behavioral inter-individual variability in principal components differed between backgrounds, individual scores for each principal component were compared between backgrounds by one-way ANOVA, followed by Tukey's posthoc test using the car package[47]. To reveal the most discriminating behavioral parameters contributing to the behavioral inter-individual variability in the PCA, a biplot of individuals and the five most discriminant variables was realized using the factoextra package[48]. Variance in behavioral parameters between backgrounds was compared by Levene's test using the car package[47]. Mean ± standard deviation (sd), and statistics, along with sex effects, are reported in **Tables S1-S3**.

**List of abbreviations**

ASD, Autism spectrum disorder

CC, Collaborative Cross

KO, knockout

LMT, Live Mouse Tracker

PCA, principal component analysis

Sd, standard deviation

WT, wildtype

**ACKNOWLEDGEMENTS**

Mouse breeding and care were performed at PAO, the rodent INRAE Animal Physiology Facility (https://doi.org/10.15454/1.5573896321728955E12). We acknowledge the use of ChatGPT, developed by OpenAI, for assistance with English editing during the preparation of this manuscript.

This project has received funding from the European Union's Horizon 2020 research and innovation programme under grant agreement N°851231. LPP and AD acknowledge the LabEx MabImprove (grant ANR-10-LABX-53-01) for the financial support of AD PhD's co-fund. The funders played no role in the conceptualization, design, data collection, analysis, decision to publish, or preparation of the manuscript.

**AUTHOR CONTRIBUTIONS**

AD and LPP designed experiments; AD, AN, CG, EP, and LPP performed experiments; AD, AN, EP, CG, and LPP contributed to the data collection; AD and NA contributed to data analysis; AD performed all statistical analysis; AD and LPP wrote the original drafts; LPP and SML contributed to the funding acquisition; LPP contributed to project conceptualization and supervision. All authors read and approved the final manuscript. All authors consent for publication.

**DATA AVAILABILITY**

Mean ± standard deviation, and statistics are represented in **Tables S1-3**. The dataset and R codes supporting the conclusions of this article is available in the Zenodo repository, https://zenodo.org/records/16354202.

**ADDITIONAL INFORMATION**

The authors declare no competing interests.

**FIGURE LEGENDS**

**Figure 1. Mixed B6;129 mice exhibit enhanced sociability and variability with unfamiliar mice.** In the 10 min three phases of the three-chambered test, B6;129 mice (orange) showed enhanced sociability (unfamiliar WT mouse 1 *vs* object; **A**), social novelty preference (familiar WT mouse 1 *vs* novel unfamiliar WT mouse 2; **B**) and familiar cage mate (familiar cage mates *vs* familiar WT mouse 2; **C**) preference compared to B6 mice (green). In the 10 min reciprocal social interaction, B6;129 mice showed enhanced social interaction with a background-, sex- and age-matched unfamiliar mouse (**D**) but not familiar cage mates (**E**) compared to B6 mice. Nose contact with an unfamiliar mouse in the reciprocal social interaction was positively correlated with those with an unfamiliar mouse in the sociability phase of the three-chambered test (**F**). In the 5 min social memory test (**G**), both B6;129 and B6 mice exhibited similar social recognition and habituation to the WT interactors over the four trials. In the odor habituation-dishabituation assay (**H**), B6;129 and B6 mice displayed similar neutral (O1, blossom flower) and social odor (O2 sex-matched or O3 opposite sex WT urines) recognition. Data are presented as individual data, mean ± sd (statistics, n and sex ratio in **Table S1**). Groups were compared by Kruskal-Wallis tests followed by Dunn post hoc tests, with stars indicating background effect and hash chamber effect (P = adjusted p-value), while inter-individual variability was compared by Levene's variance test. * or # $p < 0.05$; ** or ## $p < 0.01$; *** or ### $p < 0.001$; **** or #### $p < 0.0001$.

**Figure 2. Mixed B6;129 mice show improved social motivation and sex differences in the Live Mouse Tracker.** In the 10 min complex social interaction in the Live Mouse Tracker, B6 mice (green) and B6;129 mice (orange) displayed increased time following familiar cage mates (**A-B**) and in the subsequent trial with a mix of 2 unfamiliar B6 mice and 2 unfamiliar B6;129

mice while no differences in nose contacts were observed. Sex differences were identified when interacting with familiar cage mates (**B**), with females showing increased time huddling and in social approach compared to males in the B6;129 background, as well as nose-to-nose contacts in both backgrounds. Data are presented as individual data, mean ± sd (statistics, n and sex ratio in **Table S2**). Groups were compared by Kruskal-Wallis tests followed by Dunn post hoc tests, with stars indicating background effect and hash sex effect (P = adjusted p-value), while inter-individual variability was compared by Levene's variance test. * or # p < 0.05; ** or ## p < 0.01; *** or ### p < 0.001.

**Figure 3. Mixed B6;129 background mice display increased self-grooming.** In the 10 min motor stereotypies (**A**), B6 mice (green) displayed enhanced number of digging episodes, and no difference in the time spent digging or head shake episodes, while B6;129 mice (orange) showed increased time in self-grooming. No difference in the marble burying test (**B**) or in the Y maze alternation pattern (**C**) was observed between B6 mice and B6;129 mice. Data are presented as individual data, mean ± sd (statistics, n and sex ratio in **Table S1**). Groups were compared by Kruskal-Wallis tests followed by Dunn post hoc tests, with stars indicating background effect (P = adjusted p-value), while inter-individual variability was compared by Levene's variance test. * p < 0.05; ** p < 0.01; *** p < 0.001.

**Figure 4. Inbred B6 mice display increased anxious-like behaviors.** In the spatial object location (**A**) and novel object (**B**) recognition memory tests, B6 mice (green) and B6;129 mice (orange) performed similarly, as well travelled the same distance in the open field (**C**) for 10 min. No difference in anxious-like behavior was observed in the elevated plus maze (**D**) or time in the center in the open field arena (**E**) for both backgrounds, while B6 mice showed increased

latency to feed in the novelty-suppressed feeding tests (**F**), compared to B6;129 mice. Data are presented as individual data, mean ± sd (statistics, n and sex ratio in **Table S1**). Groups were compared by Kruskal-Wallis tests followed by Dunn post hoc tests, with stars indicating background effect (P = adjusted p-value), while inter-individual variability was compared by Levene's variance test. * $p < 0.05$; ** $p < 0.01$; *** $p < 0.001$.

**Figure 5. Sociability and motor stereotypies discriminate between backgrounds.** Based on the behavioral phenotypes of B6 mice (green) and B6;129 mice (orange) in the reciprocal social interaction, three-chambered, motor stereotypies, Y-maze, open field, and novelty-suppressed feeding tests (**A**), principal component analysis revealed higher variability in B6;129 background in the first principal component (left), confirmed by a significant increase of the scores for the first principal component in B6;129 compared to B6 background (center). The most contributing behavioral parameters to this variability in B6;129 background were sociability and stereotyped behavior parameters (right). Based on the behavioral phenotypes of B6 (green) and B6;129 mice (orange) in the Live Mouse Tracker interacting with cage mates or unfamiliar age- and sex-matched B6 and B6;129 mice (**B**), the principal component analysis revealed increased variability of the B6;129 mice in the second principal component (left), evidenced by an increase in the scores for the second component (center). The most discriminating parameters contributing to this variability in the B6;129 background were social interaction parameters (right). Data are presented as individual data, mean ± sd (**Tables S1-2**). Groups were compared by the one-way ANOVA followed by Tukey's posthoc tests, with stars indicating the background effect. * $p < 0.05$; ** $p < 0.01$; *** $p < 0.001$.

**Figure 6. Mixed B6;129 background enhances sociability and self-grooming in WT, but not in chronically isolated mice.** In the reciprocal social interaction test, chronically isolated mice (purple) revealed significant differences compared to WT controls (gray) only in the B6 background, while chronic isolation increased variance in both backgrounds. Social interaction was significantly increased in both WT control and isolated mice from the B6;129 compared to B6 background (**A**). In the sociability phase of the three-chambered test, sociability preference was observed for both chronically isolated mice and WT control, while this preference was significantly higher in B6;129 background with a higher variance (**B**). In the motor stereotypy test, there was no difference in the time spent digging or self-grooming between WT and chronically isolated mice, while both WT and isolated mice in B6 background showed increased time spent digging and number of head shakes, while increase in time spent self-grooming was observed in B6;129 WT mice compared to B6 conspecifics (**C**). Data are presented as individual data, mean ± sd (statistics, n and sex ratio in Table S3). Groups were compared by Kruskal-Wallis tests followed by Dunn post hoc tests, with stars indicating background effect, hash housing effect, and dollar chamber effect (P = adjusted p-value), while the inter-individual variability was compared by Levene's variance test. *, # or $ p < 0.05; **, ## or $$ p < 0.01; ***, ### or $$$ p < 0.001; ****, #### or $$$$ p < 0.0001.